\documentclass[sigconf]{acmart}
\usepackage{enumitem}

\usepackage{color, colortbl}
\usepackage[first=0,last=9]{lcg}
	
\definecolor{LightCyan}{rgb}{0.7,0.7,0.7}

\usepackage{xcolor}

\usepackage{algorithm}
\usepackage[noend]{algpseudocode}

\AtBeginDocument{%
  \providecommand\BibTeX{{%
    \normalfont B\kern-0.5em{\scshape i\kern-0.25em b}\kern-0.8em\TeX}}}

\begin{document}

\title[Preserving Individuality while following the Crowd]{Preserving Individuality while following the Crowd: Understanding the role of User Taste and Crowd Wisdom in Online Product Rating Prediction}

\author{Liang Wang, Shubham Jain, Yingtong Dou, Junpeng Wang, \\Chin-Chia Michael Yeh, Yujie Fan, Prince Aboagye, Yan Zheng, \\Xin Dai, Zhongfang Zhuang, Uday Singh Saini, and Wei Zhang}
\affiliation{%
  \institution{Visa Research}
  \streetaddress{900 Metro Center Blvd Foster City, CA, 94404-2775 }
  \city{Foster City}
  \state{CA}
  \country{USA}
  \postcode{94404}}
\email{{liawang, shubhjai, yidou, junpenwa, miyeh, yufan, priaboag, yazheng, xidai, zzhuang, udasaini, wzhan}@visa.com}

\renewcommand{\shortauthors}{Liang Wang et al.}


\begin{abstract}
Numerous algorithms have been developed for online product rating prediction, but the specific influence of user and product information in determining the final prediction score remains largely unexplored. Existing research often relies on narrowly defined data settings, which overlooks real-world challenges such as the cold-start problem, cross-category information utilization, and scalability and deployment issues. To delve deeper into these aspects, and particularly to uncover the roles of individual user taste and collective wisdom, we propose a unique and practical approach that emphasizes historical ratings at both the user and product levels, encapsulated using a continuously updated \textit{dynamic tree representation}. This representation effectively captures the temporal dynamics of users and products, leverages user information across product categories, and provides a natural solution to the cold-start problem. Furthermore, we have developed an efficient data processing strategy that makes this approach highly scalable and easily deployable. Comprehensive experiments in real industry settings demonstrate the effectiveness of our approach. Notably, our findings reveal that individual taste dominates over collective wisdom in online product rating prediction, a perspective that contrasts with the commonly observed ``wisdom of the crowd'' phenomenon in other domains. This dominance of individual user taste is consistent across various model types, including the boosting tree model, recurrent neural network (RNN), and transformer-based architectures. This observation holds true across the overall population, within individual product categories, and in cold-start scenarios. Our findings underscore the significance of individual user tastes in the context of online product rating prediction and the robustness of our approach across different model architectures.

\end{abstract}




\maketitle

\vspace{-0.05in}
\section{Introduction}
\label{section:Introduction}
The accurate prediction of online product ratings holds immense importance because it influences consumer purchasing decisions and provides businesses with the necessary insights to improve their products ~\cite{mcauley2013hidden,schoenmueller2020polarity,xia2021exploiting,li2008self}. Central to this prediction is a two-fold focus: the product itself and the individual user. Grasping the unique contributions of these two elements, particularly their \textit{evolving dynamics over time} both independently and collectively, is crucial for the creation of accurate rating prediction systems.

The need for such accuracy has led to the evolution of numerous innovative algorithms for rating predictions. These extend from early latent factor models~\cite{mcauley2013amateurs,xia2021exploiting,koren2009collaborative} to more recent advancements in deep learning models ~\cite{elkahky2015multi,wang2019recurrent,wu2017recurrent,yu2019adaptive,zhou2019deep,lin2022dual,ma2020temporal,he2017neural,kang2023llms}. While there is consensus on the importance of both user and product information in rating prediction, the specific roles each factor plays in the final prediction score are yet to be clearly understood.

Furthermore, much of the existing research is often based on small or pruned datasets and includes unrealistic assumptions, such as requiring a minimum number of reviews for each user and product, or being confined to specific product categories ~\cite{wang2019recurrent,kang2018self,li2020time,liu2023chatgpt,zhang2023fata,yu2019adaptive,lin2022dual}. These narrowly defined data settings can result in overlooked challenges, as outlined below:
\begin{itemize}[leftmargin=*]
\item \textbf{Artificial elimination of the cold-start problem.} The cold-start problem for users and products \cite{barjasteh2016cold,bernardi2015continuous} is a prevalent issue in real production systems. The magnitude of this problem is underscored by the product review dataset~\cite{ni2019justifying} from a major online retail platform, which is the dataset used in this paper. Out of 230,139,802 reviews contributed by 43,249,276 users across 14,894,121 products, 43.97\% of users have only provided a single review, and 37.89\% of products have received just one review. However, when using a dataset that requires at least a certain number of reviews for each user and product, such as the 5-core subset of the product review dataset~\cite{ni2019justifying}, the cold-start problem is artificially mitigated. This approach creates a skewed representation of the actual scenario because the data used to train machine learning models does not accurately mirror the broader population that the models are intended to serve.

\item \textbf{Overlooking cross-category information.} Real-world production systems frequently serve multiple product categories~\cite{ma2020temporal}. For example, the product review dataset~\cite{ni2019justifying} covers 29 product categories, spanning from highly popular categories like ``Books'' to less frequented ones such as ``Magazine Subscriptions''. It's common for a user to purchase from various product categories and provide ratings for different products. Research limited to a single product category risks neglecting valuable user information from other categories. Our experiments suggest that this cross-category information can significantly boost rating prediction performance.

\sloppy
\item \textbf{Disregarding scalability and deployment issues.} Working with smaller, pruned datasets can often mask real-world challenges related to scalability and deployment. These issues generally only become evident during the deployment stage. A model that performs well on a size-limited dataset may not necessarily maintain this performance when dealing with larger volumes of data~\cite{breck2017ml}. While the sophistication of the model is important, the way it is utilized can be even more crucial~\cite{ma2020temporal,jannach2020deep,steck2021deep,ferrari2019we}.

\end{itemize}

In this paper, we present a unique and practical approach to examine the individual and collective influences of users and products on rating prediction, without sidestepping the previously mentioned challenges. Our focus is on historical ratings at both the user and product levels, encapsulated through a \textit{dynamic tree representation}. This representation is continuously updated at each time point, effectively capturing the temporal dynamics of users and products. Moreover, this tree representation leverages user information across product categories, providing a natural solution to the challenging cold-start problem. We have developed an efficient data processing strategy that makes this approach highly scalable and easily deployable.

We conduct comprehensive experiments, demonstrating the effectiveness of our approach in a real industry data setting. Our findings notably challenge conventional wisdom observed in various problem-solving domains, where collective judgments across individuals often surpass the accuracy of individual judgments ~\cite{davis2014crowd, simoiu2019studying}. This phenomenon, commonly referred to as the ``wisdom of the crowd'', led us to hypothesize that it might also apply to product rating prediction, with aggregate user ratings for a product potentially yielding more precise predictions than a single individual's rating. However, our research indicate otherwise: \textit{the
individual user’s tastes dominates over collective wisdom in online product rating prediction}. This dominance is not only evident in the overall population but also within individual product categories and it also persists in cold-start scenarios, including user cold-start portfolios and product cold-start portfolios. The dominance of individual user tastes is consistently demonstrated across various model types, including the boosting tree model, RNN, and transformer-based architectures. Our findings highlight the significance of individual tastes and have important implications which could potentially inform future strategies for product development, marketing, and customer engagement.

\section{Related Work}
\label{section:related_work}

Numerous studies have demonstrated the effectiveness of integrating user and product information to enhance the performance of rating prediction and recommendation models. For instance, Elkahky et al.~\cite{elkahky2015multi} proposed a multi-view deep neural network model that utilizes both product features and user characteristics. Similarly, Lin et al ~\cite{lin2022dual} proposed a dual contrastive network for modeling both user sequences and product sequences for sequential recommendation. In contrast, several studies have modeled user preferences and product behaviors separately. At the user level, McAuley and Leskovec~\cite{mcauley2013amateurs} proposed a latent factor model to represent the evolution of user experience over time. At the product level, Jing et al.~\cite{jing2023capturing} presented a popularity-aware recommender framework to predict product popularity trends. Despite these advancements, the exact influence of each component on the prediction process remains unclear.

The temporal dynamics of consumer preferences and product popularity also play a crucial role in rating predictions. Simple time features have led to significant performance improvements in real-world recommender systems \cite{covington2016deep,steck2021deep}. Early representative works include using time decay functions to weigh more on recent instances~\cite{ding2005time,ji2020re,anelli2019local} and modeling user and product evolution using latent factor models ~\cite{koren2009collaborative}. Recent research has primarily used deep learning models to capture temporal user and product behavior ~\cite{wu2017recurrent,wang2019recurrent,ma2020temporal,yu2019adaptive,zhou2019deep,hidasi2015session,fan2021continuous,zhang2023fata,li2020time,ren2019lifelong,chen2019dynamic,xiang2010temporal}. A key challenge is how to represent users and products at each time step in a deep learning model. A common approach is using one-hot encoding, a technique employed in most existing deep learning models. However, this approach results in high-dimensional and sparse data. An alternative method is to use a word2vec-style technique~\cite{mikolov2013efficient} to create denser representations for both users and products \cite{wang2019recurrent}. Yet, 
updating these dense vectors at each prediction time step can be impractical when dealing with extensive datasets. Instead, a periodic batch processing (for example, monthly) and incremental pre-training for these dense vectors have been pursued~\cite{zhang2021transaction}. This causes information delay as the most recent user and product behavior cannot be timely incorporated into the model. Our proposed dynamic tree representations, along with their efficient implementation, 
provide a valuable complement for representing users and products in a deep learning model because they can be updated on the fly in much shorter time intervals. This strategy has been successfully integrated into a large-scale RNN model~\cite{zhang2021transaction}.

Addressing the cold-start problem, involving both product and user cold-start scenarios, is another crucial research area \cite{barjasteh2016cold,schein2002methods,saveski2014item,volkovs2017dropoutnet}. While strategies often use auxiliary information alongside rating data to improve predictions for new users and products, the challenge of effectively capturing the temporal dynamics of cold-start users and products remains~\cite{bernardi2015continuous}. 

Lastly, few studies have fully utilized the product review dataset to explore the benefits of leveraging information across product categories. Two notable exceptions are the works outlined in ~\cite{mcauley2013hidden,zheng2017joint}, which employed an early version of the product review dataset~\cite{ni2019justifying}. However, this version is significantly smaller than the current version of the dataset used in our study.

\begin{figure*}[h]\small
  \centering
  \includegraphics[width=0.9\linewidth]{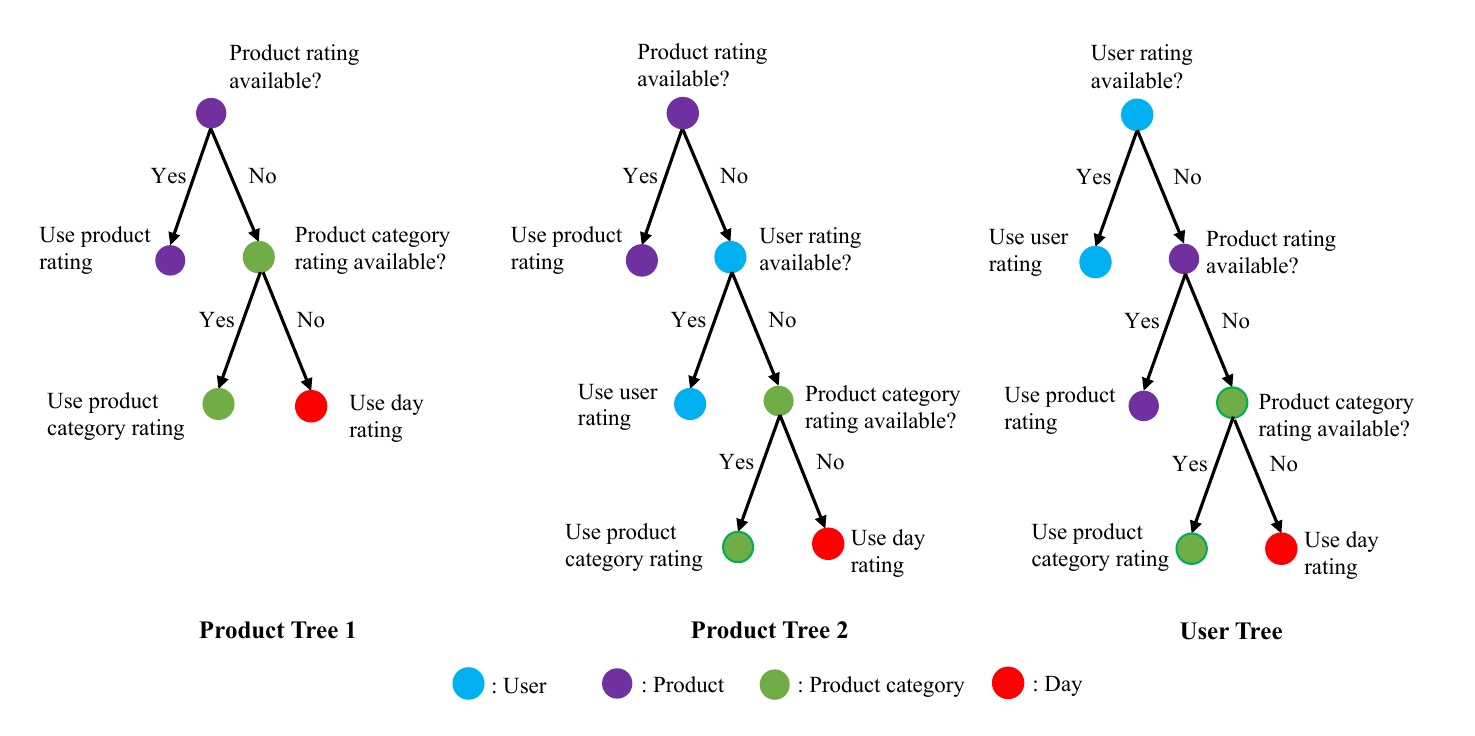}
  \vspace{-0.2in}
  \caption{``Product Tree 1'' and ``Product Tree 2'' are designed to encapsulate historical product ratings, reflecting the influence of crowd wisdom. On the other hand, the ``User Tree'' concentrates on historical user ratings, signifying individual influences. The key distinction between the two product trees lies in ``Product Tree 2''s consideration of user ratings across all product categories when ratings for the current product are not available. These trees are continuously updated at each time point $t$, with varying lengths of look-back windows, to capture the temporal dynamics of both users and products.}
  \label{fig:DiagramTreeRepresentation}
\end{figure*}

\section{Methodology}
\label{section:methodology}
In this section, we introduce our approach to elucidate the roles of individual user taste and crowd wisdom in predicting the rating that user $u$ would assign to product $p$ on a particular day $t$\footnote{Although the product review dataset in ~\cite{ni2019justifying} provides Unix timestamps for each review, these timestamps essentially represent the day of the review.}. We present the concept of \textit{dynamic tree representations}, a method designed to capture the temporal dynamics of users and products, tackle the challenging cold-start problem, and leverage cross-product category information for role discovery. We discuss different settings where dynamic tree representations are used to uncover the roles of individual user tastes and crowd wisdom. Additionally, we outline an efficient data processing pipeline that enhances the scalability and deployability of our approach.

Using historical ratings given by a user and received by a product not only offers a direct reflection of past rating behaviors, but also captures the evolving patterns over time. While the incorporation of more features, such as review text~\cite{wu2017joint,zheng2017joint,mcauley2013hidden}, could potentially enhance model performance, we intentionally limit our focus to historical ratings. This is to maintain a clear focus on our main objective and avoid deviation that could complicate the model unnecessarily~\cite{sachdeva2020useful}.

\vspace{-0.1in}
\subsection{Representation of Historic Ratings}
\label{subsection: representation}

Our goal is to understand both the individual and collective influences of users and products in online product rating prediction, based on their historical ratings. To achieve this, we require a mechanism that can effectively represent these historical ratings. We propose a simple yet efficient tree structure for this purpose, as illustrated in Figure ~\ref{fig:DiagramTreeRepresentation}. This structure comprises three distinct trees: \textbf{Product Tree 1}, \textbf{Product Tree 2}, and \textbf{User Tree}, each serving a unique role:

\begin{itemize}[leftmargin=*]
\item\textbf{\underline{Product Tree 1}} concentrates on the product's rating history. It first checks if the product has received any ratings from users (which may or may not include the current user) within the $L$ days prior to $t$, where $L$ is a \textit{look-back window}, indicating the length of the historical ratings period we should use. If the product has been rated, the tree yields its average rating. If not, it examines whether the product belongs to a rated product category. In this hierarchy, a product category acts as a parent to a product (for instance, ``Book'' is a product category, encompassing all types of books). If the product belongs to a rated category, the tree outputs the average rating for that category. If such a category doesn't exist, the tree defaults to the average rating from all reviews over the past $L$ days, excluding the ratings from the current day $t$. \textit{This tree structure aims to understand the influence of crowd wisdom in online product rating prediction}.

\item\textbf{\underline{Product Tree 2}} follows a similar process as \textbf{Product Tree 1}, but with a notable variation. If the current product hasn't received any ratings over the past $L$ days preceding $t$, the tree checks if the current user has rated any products across all categories during the same period. If so, it provides the user's average rating. This characteristic of \textbf{Product Tree 2} helps to alleviate the product cold start problem. Even when a user rates a new product that has not yet received any reviews, the fact that the user may have already rated numerous other products can provide valuable information for rating the current product. While it is possible that both the user and the product are completely new, such scenarios are extremely rare\footnote{In the product review dataset~\cite{ni2019justifying}, there are only 69,507 cases where a product has never been reviewed before, and a user gives a rating for the first time.}. \textit{This tree structure also aims to understand the influence of crowd wisdom in online product rating prediction}.

\item\textbf{\underline{User Tree}} concentrates on the user's rating history. It first checks if the user has rated any products across all categories in the $L$ days prior to $t$. If so, it provides the user's average rating. If not, it follows the same steps as \textbf{Product Tree 1}. \textit{This tree structure aims to understand the influence of individual user preference in online product rating prediction}.
\end{itemize}

These trees are dynamic and continuously updated at each time point $t$. The parameter $L$ directs our focus towards either short-term or long-term behaviors of users and products. A smaller $L$ enables us to capture more immediate behaviors, such as sudden changes in user preferences or product ratings. Conversely, a larger $L$ allows us to grasp long-term trends, helping us understand more persistent behaviors and stable patterns. The true power of our dynamic tree representations emerges when we utilize both small and large $L$ values. This approach allows us to capture the immediate fluctuations in user behavior and product ratings while still maintaining a grasp on long-term trends, thereby revealing a comprehensive view of the behavior of users and products. Further elaboration on this point will be provided in Section ~\ref{subsection:dissection_analysis}, where we conduct a \textit{dissection} analysis on trees with varying look-back window sizes. This analysis will underscore the importance of incorporating trees with varying look-back window sizes in shedding light on the roles of users and products in online product rating prediction.

Using trees with varying window sizes $L$ offers an advantage over the commonly used \textit{time decay} strategy which weighs more on recent user-product interactions~\cite{ding2005time,ji2020re,anelli2019local,jing2023capturing} because it covers not only recent user-product interactions but also all-time user-product interactions. Indeed, extensive experiments in ~\cite{koren2009collaborative} find that the best prediction performance is actually reached when there is no decay at all on instances. The reason is that despite users changing their tastes and rating scale over time, much of the old preferences may still persist.  When only considering the most recent interactions (short-term temporal dynamics), we can only obtain a partial view of users and products. It is the combination of short-term and long-term temporal dynamics that uncovers the full picture and roles of users and products~\cite{song2016multi,lai2018modeling,ren2019lifelong,chen2019dynamic,yu2019adaptive}.


Moreover, these tree representations offer a natural solution to the cold-start problem. When a user $u$ has not posted any reviews before the current time $t$ (prediction time), we encounter the \textit{user cold-start problem}\footnote{To be more accurate, we should refer to this as either a missing value or cold-start problem. Within a given look-back window $L$, if the ratings, either from users or products, are not available, it does not necessarily indicate a cold-start problem unless $L$ represents the entire lifespan. Nonetheless, our approach handles both missing values and cold-start scenarios within a unified framework.}. Conversely, when a product $p$ has not received any reviews prior to the current time $t$, we face the \textit{product cold-start problem}. In such cases, we can rely on the appropriate tree representations to predict ratings. For the user cold-start problem, either \textbf{Product Tree 1} or \textbf{Product Tree 2} can be used, and under these circumstances, both trees become identical. For the product cold-start problem, either the ~\textbf{User Tree} or ~\textbf{Product Tree 2} can be employed, in which case, these two trees become identical. This dynamic approach, where \textbf{Product Tree 1}, \textbf{Product Tree 2}, and \textbf{User Tree} are continually updated with varying $L$ at every time point $t$, helps to alleviate the continuous cold-start problem~\cite{bernardi2015continuous}.

\subsection{Using Dynamic Tree Representations for Role Discovery}
\label{subsection: using_tree_representations}

Outputs from instances of \textbf{Product Tree 1}, \textbf{Product Tree 2}, and \textbf{User Tree} can serve as features to a downstream model. Here, an instance corresponds to a specific look-back window size, $L$. A smaller $L$, like 7 days, captures short-term behaviors, while a larger one, such as the lifetime span, targets long-term behaviors. 
In this study, we set $L$ to 7d, 30d, 90d, 1y, 3y, 5y, and the full lifespan.

More specifically, we establish the following settings for role discovery, each with a unique purpose:

\begin{itemize}[leftmargin=*]
\sloppy
\item \textbf{\underline{Setting $\mathbf{S1}$ (Crowd Wisdom)}}. This setting includes outputs from \textit{seven} instances of \textbf{Product Tree 1} as inputs to a model. Each instance corresponds to a different look-back window size ($L=7d, 30d, 90d, 1y, 3y, 5y,$\textit{lifespan}), allowing the model to capture both short-term and long-term temporal dynamics of a product. The primary purpose of this setting is to analyze the impact of collective intelligence, or crowd wisdom, in online product rating prediction.

\item \textbf{\underline{Setting $\mathbf{S2}$ (Crowd Wisdom)}}. This setting utilizes outputs from \textit{seven} instances of \textbf{Product Tree 2} as inputs to a model. Similar to $\mathbf{S1}$, its purpose is to analyze the influence of crowd wisdom in online product rating prediction. However, it differs in a key aspect: \textbf{Product Tree 2} draws upon user rating information from other product categories when ratings for the current product are not available within a given $L$. This setting helps to address product cold start scenarios.

\item \textbf{\underline{Setting $\mathbf{S3}$ (Individual Taste)}}. This setting uses outputs from \textit{seven} instances of \textbf{User Tree} as inputs to a model, focusing on analyzing the impact of individual user tastes in online product rating prediction.

\item \textbf{\underline{Setting $\mathbf{S4}$ (Crowd Wisdom + Individual Taste)}}. This setting incorporates outputs from all the previously mentioned instances of \textbf{Product Tree 1}, \textbf{Product Tree 2}, and \textbf{User Tree} (a total of 21) as inputs to a model, aiming to analyze the combined impact of crowd wisdom and individual user tastes in online product rating prediction.

\end{itemize}

All models used in subsequent experiments (Section ~\ref{sec:Experiments}), including the boosting tree model, RNN, and transformer-based architectures, will be based on these four settings.

\subsection{Efficient Data Processing}
\label{subsection:data_processing}
Our proposed method involves a retrospective analysis of historical data, with various look-back window sizes for each user, product, and product category. In a production setting, even though the model parameters might remain constant post-deployment, the model inputs, especially the representations derived from historical data for the aforementioned entities, need continuous updates at each time step to capture temporal dynamics. Consequently, efficient data processing becomes essential for successfully implementing a real-world model~\cite{ma2020temporal,steck2021deep}. These challenges are often not encountered when working with smaller or pruned datasets, which are typically the focus of most existing research. To address these challenges, we've designed an efficient data processing pipeline that includes several key steps, as summarized in Algorithm 1.

\begin{algorithm}
\caption{Data Processing Pipeline}
\begin{algorithmic}[1]

\Procedure{PartitionData}{\textit{raw\_data}}
    \State Store raw data in HDFS and partition it by day
\EndProcedure

\Procedure{DailyAggregation}{\textit{partitioned\_data}}
    \State Aggregate daily statistics, e.g., number of reviews and sum of ratings
\EndProcedure

\Procedure{ComputeAverageRatings}{\textit{aggregated\_data, window\_sizes}}
    \State Compute average ratings based on daily-aggregated statistics with varying sizes of look-back windows
\EndProcedure

\Procedure{BuildTrees}{\textit{average\_ratings, window\_sizes}}
    \State Build User Trees and Product Trees with varying sizes of look-back windows
\EndProcedure

\Procedure{GenerateModelingData}{\textit{label, user\_trees, product\_trees}}
    \State Generate modeling data at the instance (review) level by appending the prediction label to trees
\EndProcedure

\end{algorithmic}
\end{algorithm}

\begin{figure}[h]\small
  \centering
  \includegraphics[width=60mm, scale=0.1]
  {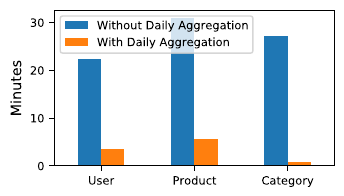}
  \vspace{-0.1in}
  \caption{Time needed with and without daily aggregation.}
  \label{fig:TimeDailyAggregation}
\end{figure}

Among these steps, the second one, \textit{Daily Aggregation}, is crucial. It significantly reduces the overall processing time since a user/product can give/receive multiple reviews in a single day. For instance, collecting user ratings and constructing seven instances of \textbf{User Tree} (that is, setting $\mathbf{S3}$) for just one day of data in the product review dataset~\cite{ni2019justifying}, such as June 1st, 2018, takes approximately 22 minutes without the \textit{Daily Aggregation} step. Extrapolating this to include an entire year's worth of data would require over 5 days for a single Hive job ~\cite{thusoo2009hive} in the Hadoop Distributed File System (HDFS)~\cite{shvachko2010hadoop}. However, with the \textit{Daily Aggregation} step, this time is reduced to under four minutes for the same one day of data and to less than one day for an entire year's worth of data. This significantly accelerates our ability to process and analyze large volumes of data, making our approach highly scalable and easily deployable in a production environment. This approach has been incorporated into a large-scale RNN model processing billions of transaction events~\cite{zhang2021transaction}. Figure ~\ref{fig:TimeDailyAggregation} provides a comparison illustrating the benefits of using \textit{Daily Aggregation} compared to not using it on one-day data (June 1st, 2018).

\section{Experiments}
\label{sec:Experiments}

In this section, we utilize our proposed approach to conduct experiments on a large-scale industrial dataset. Our objective is to elucidate the roles of individual user tastes and crowd wisdom in online product rating prediction. Specifically, our experiments are designed to answer the following research questions:

\sloppy
\textbf{RQ1}: How do individual user tastes and crowd wisdom individually and jointly influence online product rating prediction? \\
\indent\textbf{RQ2}: How do these influences vary across different portfolios, including \textit{warm-start} and \textit{cold-start} scenarios?\\
\indent\textbf{RQ3}: Why is it crucial to incorporate trees with various look-back window sizes when examining these influences? \\ 
\indent\textbf{RQ4}: Is it more effective to use a single model trained on data from all product categories or multiple models each trained for a specific product category? \\
\indent\textbf{RQ5}: Do model types have an impact on the findings? \\

\subsection{Dataset and Data Splitting}
\label{dataset}
We utilize the full product review dataset\footnote{\url{https://nijianmo.github.io/amazon/index.html}}~\cite{ni2019justifying}, which consists of 230,139,802 reviews from 43,249,276 users, spanning 14,894,121 unique products across 29 distinct categories. The dataset covers the period from May 1996 to October 2018, providing an extensive view of user-product interactions. Ratings in the dataset range from 1 to 5, with higher values indicating stronger preferences. 

In contrast to the common practice of random data splitting, we implement a real-world production scenario for our data splitting. In this setup, models are trained and validated using data up to a certain date, and then tested on unseen future data, also known as out-of-time data. This strategy helps us address information leakage, an issue that is prevalent in many research papers and data mining competitions, as highlighted in several studies~\cite{wang2022learning, kaufman2012leakage,ji2023critical}. Specifically, we designate data from January 2017 to October 2018 as out-of-time testing data, while data from 2016 is used for model validation, and data prior to 2016 for model training. This setup allows us to evaluate the model's performance on previously unseen data, reflecting real-world scenarios.

\subsection{Labeling Logic and Evaluation Metric}
\label{labeling_logic_and_metric}

Considering that most users simplify their decision-making to a binary pattern of either liking or disliking a product~\cite{herlocker1999algorithmic,schoenmueller2020polarity}, the difference between a prediction of 1.5 and 2.5 becomes irrelevant if the user is only interested in products rated 4 or higher. Therefore, we adopt the labeling approach used in prior research~\cite{schein2002methods,xia2021exploiting,zhang2023fata,bao2023tallrec}, and treat the rating prediction task as a binary classification problem. In this framework, ratings above 3 stars are classified as 1 (indicating likeness), and all others as 0 (indicating disapproval).

To evaluate the predictive performance of our models, we utilize the area under the receiver operating characteristic (ROC) curve (AUC), a metric particularly suited for assessing binary decision models~\cite{herlocker1999algorithmic}. 


\subsection{Models Used}
\label{subsection:models_used}
We conduct experiments utilizing three types of models: LightGBM~\cite{ke2017lightgbm}, RNN~\cite{cho2014learning,hochreiter1997long}, and TabTransformer~\cite{huang2020tabtransformer}. LightGBM was selected due to its demonstrated effectiveness in various Kaggle competitions and industry applications, as well as its superior performance in numerous benchmark comparisons~\cite{jannach2020deep,grinsztajn2022tree,shwartz2022tabular}. The RNN model was chosen in light of the sequential nature of our data and the presence of unique identifiers (reviewer ID) within the data~\cite{lipton2015critical}. Our RNN is based on a production-oriented setting as detailed in ~\cite{zhang2021transaction}. Lastly, we incorporated TabTransformer~\cite{huang2020tabtransformer}, a pioneering work in applying Transformer architectures to tabular data, which is often used by researchers as both an inspiration and a performance benchmark~\cite{shwartz2022tabular,grinsztajn2022tree,somepalli2021saint}. The implementation details for these three types of models can be found in the Appendix \ref{appendix_A}. The findings across these three types of models are consistent, and hence, the experimental results reported in the subsequent sections are primarily based on the LightGBM model. However, in Section ~\ref{subsection:does_model_type}, we provide a performance comparison of the three types of models on the overall population, and in Appendix \ref{appendix_B}, a detailed performance comparison across 29 individual product categories. All the models are trained using features based on the four settings, $\mathbf{S1}$, $\mathbf{S2}$, $\mathbf{S3}$, and $\mathbf{S4}$, as outlined in Section~\ref{subsection: using_tree_representations}.






\subsection{RQ1. Individual and Joint Roles of Products and Users in Overall Population and Individual Product Categories}
\label{subsection:individual_and_joint_role}
Table \ref{table: IndividualRole} presents AUC values obtained from the LightGBM model under these four settings for both the overall product category and the 29 individual categories on the testing dataset. The table is organized based on the number of reviews in the entire dataset, with the ``Books'' category having the highest number of reviews and ``Magazine Subscriptions'' having the fewest. Please also refer to Figure ~\ref{fig:AUC_lightGBM_appendix} in Appendix \ref{appendix_B} for plots related to these four settings.

\begin{table}[ht]
    \caption {Overall and individual AUC values across 29 product categories under four different settings. The five rows highlighted in color represent the top five product categories that align closely with personal tastes.}
    \vspace{-0.1in}
    \centering
    \resizebox{\columnwidth}{!}{%
    \begin{tabular}{|l|l|l|l|l|l|}
    \hline
    Index & Category & $\mathbf{S1}$ & $\mathbf{S2}$ & $\mathbf{S3}$ & $\mathbf{S4}$ \\ \hline
        0 & Overall & 0.6965 & 0.7025 & 0.7297 & 0.7493 \\ \hline
        \rowcolor{green}
        1 & Books & 0.6892 & 0.7053 & 0.7572 & 0.7778 \\ 
        \hline
        2 & Clothing\_and\_Shoes& 0.6516 & 0.6565 & 0.6846 & 0.7029 \\ \hline
        3 & Home\_and\_Kitchen & 0.6901 & 0.6940 & 0.7193 & 0.7404 \\ \hline
        4 & Electronics & 0.6899 & 0.6925 & 0.7230 & 0.7398 \\ \hline
        5 & Sports\_and\_Outdoors & 0.6779 & 0.6829 & 0.7088 & 0.7303 \\ \hline
        6 & Cell\_Phones\_and\_Accs & 0.6837 & 0.6863 & 0.7151 & 0.7280 \\ \hline
        7 & Tools\_and\_Home\_Imprv & 0.6880 & 0.6931 & 0.7226 & 0.7457 \\ \hline
        8 & Movies\_and\_TV & 0.7245 & 0.7298 & 0.7405 & 0.7703 \\ \hline
        9 & Toys\_and\_Games & 0.7210 & 0.7265 & 0.7375 & 0.7639 \\ \hline
        10 & Automotive & 0.6890 & 0.6975 & 0.7114 & 0.7388 \\ \hline
        11 & Pet\_Supplies & 0.6906 & 0.6924 & 0.7146 & 0.7333 \\ \hline
        \rowcolor{green}
        12 & Kindle\_Store & 0.6706 & 0.7089 & 0.7602 & 0.7842 \\ \hline
        13 & Office\_Products & 0.6948 & 0.6992 & 0.7291 & 0.7511 \\ \hline
        14 & Patio\_Lawn\_Garden & 0.7125 & 0.7163 & 0.7350 & 0.7596 \\ \hline
        15 & Grocery\_and\_Gourmet & 0.6702 & 0.6748 & 0.7059 & 0.7277 \\ \hline
        \rowcolor{green}
        16 & CDs\_and\_Vinyl & 0.6167 & 0.6437 & 0.6850 & 0.7093 \\ \hline
        17 & Arts\_Crafts\_Sewing & 0.6813 & 0.6921 & 0.7175 & 0.7441 \\ \hline
        18 & Video\_Games & 0.6889 & 0.6925 & 0.7189 & 0.7335 \\ \hline
        19 & Industrial\_and\_Sci & 0.6753 & 0.6845 & 0.7153 & 0.7392 \\ \hline
        20 & Digital\_Music & 0.6446 & 0.6818 & 0.6849 & 0.7195 \\ \hline
        21 & Musical\_Instruments & 0.6843 & 0.6895 & 0.7104 & 0.7340 \\ \hline
        22 & Amazon\_Fashion & 0.6737 & 0.6842 & 0.6922 & 0.7168 \\ \hline
        23 & Appliances & 0.7003 & 0.7033 & 0.7241 & 0.7462 \\ \hline
        \rowcolor{green}
        24 & Luxury\_Beauty & 0.6227 & 0.6244 & 0.6688 & 0.6836 \\ \hline
        25 & Prime\_Pantry & 0.6505 & 0.6534 & 0.6940 & 0.7228 \\ \hline
        26 & Software & 0.6766 & 0.6815 & 0.7174 & 0.7370 \\ \hline
        27 & All\_Beauty & 0.6800 & 0.6865 & 0.7009 & 0.7236 \\ \hline
        28 & Gift\_Cards & 0.6871 & 0.6898 & 0.7167 & 0.7561 \\ \hline
        \rowcolor{green}
        29 & Magazine\_Subscriptions & 0.6610 & 0.6664 & 0.7104 & 0.7321 \\ \hline
    \end{tabular}%
    \label{table: IndividualRole}
    }
\end{table}

\textit{The results compellingly demonstrate the dominance of individual user tastes over crowd wisdom}: $\mathbf{S3}$ (which represents individual user tastes) consistently outperforms $\mathbf{S1}$ and $\mathbf{S2}$ (which represent crowd wisdom), not only in the overall population but also across all 29 categories. This dominance is particularly pronounced in product categories that align closely with personal tastes. These categories are highlighted in color, revealing fascinating patterns. For instance, when comparing $\mathbf{S3}$ and $\mathbf{S1}$, the top five categories with the largest relative AUC differences (calculated as (AUC($\mathbf{S3}$)-AUC($\mathbf{S1}$))/AUC($\mathbf{S1}$)) are ``Kindle Store'', ``CDs and Vinyl'', ``Books'', ``Magazine Subscriptions'', and ``Luxury Beauty'', showing differences of 13.36\%, 11.08\%, 9.87\%, 7.47\%, and 7.40\% respectively. Similarly, when comparing $\mathbf{S3}$ and $\mathbf{S2}$, the top five categories in terms of largest relative AUC differences (calculated as (AUC($\mathbf{S3}$)-AUC($\mathbf{S2}$))/AUC($\mathbf{S2}$)) are ``Books'', ``Kindle Store'', ``CDs and Vinyl'', ``Luxury Beauty'', and "Magazine Subscriptions", with differences of 7.36\%, 7.24\%, 7.11\%, 6.60\%, and 6.42\% respectively. 

It is interesting to note that the ``All Beauty'' category (the row with Index=27), which one might assume to have similarities with ``Luxury Beauty'' (the row with Index=24), actually exhibits a distinct pattern. Among the 29 product categories, it ranks fourth from the bottom for the largest relative AUC difference between $\mathbf{S3}$ and $\mathbf{S1}$ (3.07\%), and sixth for the largest relative AUC difference between $\mathbf{S3}$ and $\mathbf{S2}$ (2.10\%). This suggests a difference in the influence of personal preferences and crowd preferences between these categories. ``Luxury Beauty'' shows a stronger correlation with personal preferences, while ``All Beauty'' seems to align more with the broader crowd's preferences. This could be attributed to various factors, including differences in price points and accessibility\footnote{Coincidentally, there is a ``Beauty'' category in the early version of the product review dataset, which has been one of the most frequently used datasets for testing new algorithms in recommender systems. The pioneering SASRec algorithm~\cite{kang2018self} initially used the ``Beauty'' dataset, setting a benchmark that subsequent studies have used for comparison~\cite{li2020time,liu2023chatgpt,chen2022denoising}}.



These patterns suggest that individual user tastes significantly influence rating predictions, especially in categories where personal preferences are highly variable and subjective, offering valuable insights for enhancing recommendation systems and personalizing user experiences.


Regarding settings $\mathbf{S1}$ and $\mathbf{S2}$, both of which are designed to analyze crowd wisdom, $\mathbf{S2}$ consistently outperforms $\mathbf{S1}$. This superior performance can be attributed to $\mathbf{S2}$'s strategy of utilizing user rating information from various product categories when current product ratings are unavailable, a feature that $\mathbf{S1}$ does not possess.

The setting $\mathbf{S4}$, which incorporates both personal tastes and crowd wisdom, demonstrates a significant performance improvement compared to the individual models. This result suggests that while crowd wisdom may be less dominant, its combination with individual user tastes can substantially boost model performance.

\subsection{RQ2. Individual and Joint Roles of Products and Users across Different Portfolios}
\label{subsection:performance_by_portfolios}
Table \ref{table: performance_cold_start} presents AUC values obtained from the LightGBM model under four settings across different portfolios. These portfolios include warm-start scenarios, where users and products have at least one review prior to the prediction time, and cold-start scenarios, where users and products lack reviews before the prediction time. \textit{It is evident that the setting $\mathbf{S3}$, representing individual user tastes, consistently outperforms the settings $\mathbf{S1}$ and $\mathbf{S2}$, which represent crowd wisdom, across all these portfolios}.


The comparison between $\mathbf{S1}$ and $\mathbf{S2}$, both of which focus on crowd wisdom, is also noteworthy. The key difference lies in the fact that $\mathbf{S2}$ utilizes user rating information from all product categories, a feature that $\mathbf{S1}$ lacks. This inclusion leads to an enhanced performance of $\mathbf{S2}$ across all user and product portfolios compared to $\mathbf{S1}$. This advantage is most pronounced in product cold-start scenarios: when user information is present, the AUC value escalates to 0.6719, whereas it decreases to 0.5737 in its absence.

Furthermore, the last column ($\mathbf{S4}$) of Table \ref{table: performance_cold_start} underscores the importance of leveraging both individual tastes and crowd wisdom to achieve the greatest benefits.

\vspace{-0.05in}
\begin{table}[h]\small
\caption {AUC across different portfolios under four settings.}
\vspace{-0.1in}
\begin{tabular}{|l|l|l|l|l|}
\hline
Portfolio\textbackslash{}Setting & $\mathbf{S1}$ & $\mathbf{S2}$  & $\mathbf{S3}$ & $\mathbf{S4}$\\ \hline
User warm-start               & 0.6909   & 0.6977 & 0.7290 & 0.7489     \\ \hline
User cold-start               & 0.7337   & 0.7344 & 0.7361 & 0.7372     \\ \hline
Product warm-start            & 0.6985   & 0.7031 & 0.7308 & 0.7507     \\ \hline
Product cold-start            & 0.5737   & 0.6719 & 0.6731 & 0.6737     \\ \hline
\end{tabular}
\label{table: performance_cold_start}
\end{table}

\subsection{RQ3. Dissection Analysis}
\label{subsection:dissection_analysis}

Our four settings, $\mathbf{S1}$, $\mathbf{S2}$, $\mathbf{S3}$, and $\mathbf{S4}$, incorporate trees with a range of look-back window sizes ($L=7d, 30d, 90d, 1y, 3y, 5y,$\textit{lifespan}). This crucial feature allows our models to capture both short-term and long-term temporal dynamics of products and users, thereby presenting a comprehensive view of product and user behaviors.

To further illustrate the benefits and significance of integrating trees with different look-back window sizes, we conduct a dissection analysis. Figure \ref{fig:SingleVsMultiple} displays the AUC values derived from individual trees, each corresponding to a distinct look-back window size. In this scenario, the output from each tree is utilized as the prediction score for the rating.

\begin{figure}[h]\small
  \centering
  \includegraphics[width=60mm, scale=0.1]
  {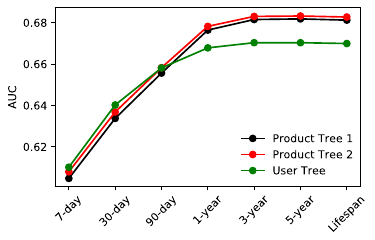}
  \vspace{-0.2in}
  \caption{AUC values of individual trees with varying look-back window sizes.}
  \label{fig:MovingWindow}
\end{figure}

The dissection analysis reveals an interesting pattern. For window sizes less than 90 days (shorter term), \textbf{User Tree}, designed to represent individual user tastes, slightly outperforms \textbf{Product Tree 1} and \textbf{Product Tree 2}, which are designed to capture crowd wisdom. However, when the window size extends beyond 90 days (longer term), the AUC of \textbf{User Tree} is significantly outperformed by the two product trees. At first glance, these findings seem to conflict with the conclusions that we have drawn from previous experiments (refer to Table ~\ref{table: IndividualRole}) which suggest that individual user tastes prevail over crowd wisdom in rating prediction. Yet, this apparent contradiction underscores the importance of using an integrated approach that incorporates all window sizes. When each tree is evaluated separately, it only provides a \textit{partial} picture of the dynamics at play. On the other hand, the integrated approach used in our settings $\mathbf{S1}$, $\mathbf{S2}$, $\mathbf{S3}$, and $\mathbf{S4}$ captures both the short-term and long-term dynamics, providing a more holistic representation of the rating behaviors of products and users. \textit{This emphasizes the importance of not relying solely on a single tree corresponding to a specific window size, but rather integrating across all trees corresponding to various window sizes to capture the \textit{full} temporal dynamics of individual user tastes and crowd wisdom}.


\subsection{RQ4. Single Model or Multiple Models?}
\label{subection:single_model_vs_multiple_models}

\begin{figure}[h]\small
  \centering
  \includegraphics[width=\columnwidth]{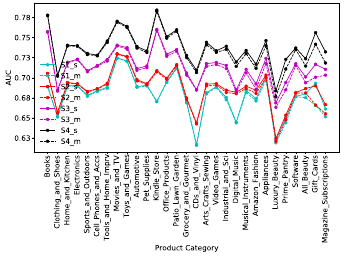}
  \vspace{-0.2in}
  \caption{This figure compares the performance of two modeling approaches: one using a single model for all categories (solid lines) and another employing 29 separate models, each for a specific product category (dotted lines). The four different colors represent four distinct settings: $\mathbf{S1}$, $\mathbf{S2}$, $\mathbf{S3}$, and $\mathbf{S4}$. The suffix "\_s" signifies the single model approach, while the suffix "\_m" indicates the multiple model approach.}
  \label{fig:SingleVsMultiple}
\vspace{-0.15in}
\end{figure}

All experiments conducted thus far have involved training one model that covers all 29 individual product categories. It could be argued that using a distinct model for each category, thereby creating a total of 29 specialized models, could potentially yield superior performance due to their increased specialization. However, our experiments challenge this proposition, as illustrated in Figure \ref{fig:SingleVsMultiple}. In this experiment, we constructed 29 LightGBM models, each tailored to a specific product category, using the same input features as the single model. This was carried out for all four settings $\mathbf{S1}$, $\mathbf{S2}$, $\mathbf{S3}$, and $\mathbf{S4}$, resulting in a total of 4 * 29 models. 
Interestingly, \textit{the single model consistently matches or surpasses the performance of the corresponding specialized models}. This performance disparity is especially noticeable for smaller product categories, underscoring the benefits of utilizing more comprehensive data and cross-product information in rating prediction. Moreover, choosing a single model over multiple specialized models offers a significant advantage for deployment, simplifying the management of the production environment by reducing the need to oversee numerous models.

\subsection{RQ5. Does Model Type Make a Difference?}
\label{subsection:does_model_type}
Until now, all the results have been obtained using the LightGBM model, under four distinct settings: $\mathbf{S1}$, $\mathbf{S2}$, $\mathbf{S3}$, and $\mathbf{S4}$. Naturally, one might question whether these results hold true for other models, such as the RNN model and the TabTransformer. \textit{Our experiments confirm the consistency of these findings across these model types, indicating that the results are model-agnostic}. Table~\ref{table: performance_model_type} displays the AUC values for the RNN and TabTransformer models, alongside the previously reported AUC values for the LightGBM model for comparison, for the entire 29 product categories. More detailed information about these models' performance on 29 individual product categories can be found in Appendix~\ref{appendix_B}. 

As observed in Table~\ref{table: performance_model_type}, although the AUC values vary across different model types and settings, the established pattern from LightGBM is reaffirmed by RNN and TabTransformer: models under setting $\mathbf{S3}$ (representing individual user tastes) consistently outperform those under settings $\mathbf{S1}$ and $\mathbf{S2}$ (representing crowd wisdom). Models under setting $\mathbf{S2}$, which leverage user rating information from various product categories when current product ratings are unavailable, consistently outperform models under setting $\mathbf{S1}$ that do not utilize such information. Furthermore, models that integrate personal tastes and crowd wisdom (setting $\mathbf{S4}$) consistently yield the most significant benefits, outperforming models that exclusively use either setting $\mathbf{S1}$, $\mathbf{S2}$, or $\mathbf{S3}$.

\vspace{-0.05in}
\begin{table}[h]\small
\caption {AUC values from different model types on entire 29 product categories.}
\vspace{-0.1in}
\begin{tabular}{|l|l|l|l|l|}
\hline
Model \textbackslash{} Setting & $\mathbf{S1}$ & $\mathbf{S2}$   & $\mathbf{S3}$ & $\mathbf{S4}$\\ \hline
LightGBM               & 0.6965   & 0.7025 & 0.7297 & 0.7493     \\ \hline
RNN               & 0.7056   & 0.7107 & 0.7286 & 0.7500     \\ \hline
TabTransformer           & 0.6935   & 0.6983 & 0.7197 & 0.7424     \\ \hline
\end{tabular}
\label{table: performance_model_type}
\end{table}


\vspace{-0.1in}
\section{Discussions}
\label{sec:discussion}

This study is primarily invested in uncovering the roles of individual user preferences and collective intelligence in online product rating prediction, rather than engaging in comparative analysis of model performance. Our approach is specifically centered on the historical ratings given by users and received by products. This not only offers a direct reflection of past rating behaviors, but also captures the evolving patterns over time. Although the inclusion of additional features may potentially improve model performance, we deliberately confine our attention to historical ratings. This is to preserve a clear focus on our primary objective and avoid deviation that could complicate the model unnecessarily~\cite{sachdeva2020useful}. However, we also acknowledge the potential benefits of feature augmentation. As part of this exploration, we conduct an experiment where we generate a comprehensive set of 136 features, including the original 21 tree features, related to users and products. These features are then used as inputs for the LightGBM, RNN, and TabTransformer models. Table \ref{table: performance_more_features} presents the AUC values of these three models for the entire 29 product categories. For comparison, also included in the table are the AUC values from these three models reported from Table ~\ref{table: performance_model_type} based on setting $\mathbf{S4}$ which includes only 21 instances of the three representations. The inclusion of a broader set of features not only boosts the AUC value of LightGBM but also improves those of RNN and TabTransformer, in contrast to models that solely utilize the 21 tree features. This enhancement is particularly substantial for TabTransformer, which sees its AUC value rise from 0.7424 to 0.7565. While deep learning models are typically recognized for their ability to discern complex patterns without extensive feature engineering, this experiment suggests that feature engineering can indeed augment the performance of these models. Additional details on features and these models' performance on individual product categories can be found in Appendix~\ref{appendix_C}.

\vspace{-0.05in}
\begin{table}[h]\small
\caption {AUC values from different model types with additional features on entire 29 product categories, compared to previous results ($\mathbf{S4}$ in Table ~\ref{table: performance_model_type}).}.

\vspace{-0.1in}
\begin{tabular}{|l|l|l|l|}
\hline
Setting /Model & LightGBM & RNN & TabTransformer \\ \hline

$\mathbf{S4}$ (21 features)   & 0.7493   & 0.7500 & 0.7424     \\ \hline
136 features   & 0.7562   & 0.7562 & 0.7565     \\ \hline

\end{tabular}
\label{table: performance_more_features}
\end{table}

Regarding the dataset used for this study, to the best of our knowledge, this is the first study that utilizes the complete product review dataset~\cite{ni2019justifying}, encompassing all 29 product categories, without modifying the original data. Although this dataset provides a comprehensive representation of user-product interactions on a large scale, its scope is limited to a single type of platform - online retail. Future work will investigate whether the findings of this study remain consistent, and the methodologies proposed here can be applied to data from diverse platforms such as movie rentals, food delivery, and travel booking.
\section{Conclusion}
\label{sec:conclusion}

In this study, we have unveiled the dominance of individual user tastes in online product rating prediction, thereby challenging the conventional ``wisdom of the crowd'' phenomenon within a large-scale industry data setting. Our findings, which hold true across various scenarios, underline the substantial role of individual user tastes. By employing a dynamic tree representation, we have successfully captured both individual and collective influences along with their temporal dynamics, without sidestepping the challenges associated with cold-start problems, cross-product category utilization, and scalability and deployment issues. These findings not only introduce a new perspective on online product rating prediction but also make a valuable contribution to the existing body of knowledge.


\bibliographystyle{ACM-Reference-Format}
\bibliography{main_paper}


\appendix
\section{Model Training Setups}
\label{appendix_A}
We utilize the LightGBM Python library\footnote{\url{https://pypi.org/project/lightgbm/}} for our LightGBM model. The hyperparameters are configured as follows: learning rate = 0.01, number of leaves = $2^8$, maximum number of bins = 255, subsample = 0.7, number of iterations = 10,000, and early stopping rounds = 100. Early stopping decisions are based on the AUC score of the validation dataset.

The RNN model is trained using the Pytorch library. Following a similar approach as described in ~\cite{kang2018self,li2020time,zhang2023fata}, each user sequence is limited to a maximum length of 10, with sequences shorter than this left-padded with a special token. The RNN model structure includes a single GRU layer with 256 hidden neurons, followed by two fully connected (FC) layers with dimensions 256 and 3, respectively. Here, 3 corresponds to the number of classes (high rating, low rating, and padding), a setting similar to that used in ~\cite{zhang2021transaction}. We apply softmax to the output and use cross-entropy loss for back-propagation. The Adam optimizer is used with a learning rate of 1e-4. Given the large size of the dataset, user sequences are converted into individual files, allowing us to avoid reading the entire data at once. For faster data reading, sequences are chunked into groups of 128 and stored in a single file, rather than 128 separate files. We then randomly select four files at a time, resulting in a batch size of 512.

Our TabTransformer setup comprises four transformer layers, each with four self-attention heads, followed by two FC layers. These FC layers adjust the transformer's output size from 32 to 64 and finally to 2. The model is implemented using the Pytorch-Widedeep Library~\cite{zaurin2023pytorch} and is trained with a batch size of 10,000, using the Adam optimizer with a learning rate of 1e-4.

\section{More Details on RQ5: Does Model Types Make a Difference?}
\label{appendix_B}

\begin{figure}[h]\small
  \centering
  \includegraphics[width=\columnwidth]{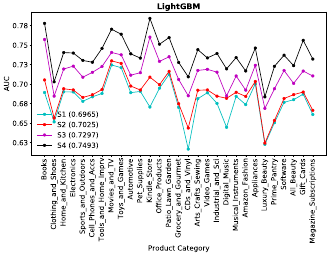}
  \vspace{-0.2in}
  \caption{This figure displays the AUC values of LightGBM models under four different settings, $\mathbf{S1}$, $\mathbf{S2}$, $\mathbf{S3}$, and $\mathbf{S4}$, across 29 individual product categories. The values in parentheses represent the AUC values for the entire product categories.}
  \label{fig:AUC_lightGBM_appendix}
\end{figure}

\begin{figure}[h]\small
  \centering
  \includegraphics[width=\columnwidth]{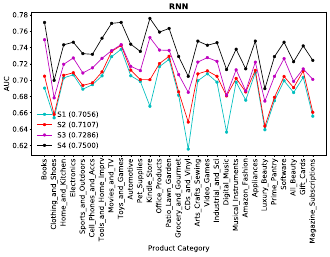}
  \vspace{-0.2in}
  \caption{This figure displays the AUC values of RNN models under four different settings, $\mathbf{S1}$, $\mathbf{S2}$, $\mathbf{S3}$, and $\mathbf{S4}$, across 29 individual product categories. The values in parentheses represent the AUC values for the entire product categories.}
  \label{fig:AUC_RNN_appendix}
\end{figure}

\begin{figure}[h]\small
  \centering
  \includegraphics[width=\columnwidth]{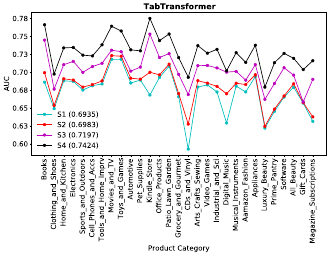}
  \vspace{-0.2in}
  \caption{This figure displays the AUC values of TabTransformer models under four different settings, $\mathbf{S1}$, $\mathbf{S2}$, $\mathbf{S3}$, and $\mathbf{S4}$, across 29 individual product categories. The values in parentheses represent the AUC values for the entire product categories.}
  \label{fig:AUC_TabTransformer_Appendix}
\end{figure}

Figures ~\ref{fig:AUC_lightGBM_appendix}, ~\ref{fig:AUC_RNN_appendix}, and ~\ref{fig:AUC_TabTransformer_Appendix} present AUC values obtained from three different model types, LightGBM, RNN, and TabTransformer, for the 29 individual categories under four distinct settings $\mathbf{S1}$, $\mathbf{S2}$, $\mathbf{S3}$, and $\mathbf{S4}$. These figures are organized based on the number of reviews in the entire dataset, with the ``Books'' category having the highest number of reviews and ``Magazine Subscriptions'' having the fewest. As observed, models under setting $\mathbf{S3}$ (representing individual user tastes) consistently outperform those under settings $\mathbf{S1}$ and $\mathbf{S2}$ (representing crowd wisdom). Models under setting $\mathbf{S2}$ also outperform those under setting $\mathbf{S1}$. It's worth noting that both $\mathbf{S1}$ and $\mathbf{S2}$ are designed to represent crowd wisdom, but $\mathbf{S2}$ utilizes user rating information from other product categories when current product ratings are unavailable, a feature that $\mathbf{S1}$ does not possess. Furthermore, models under setting $\mathbf{S4}$, which incorporate both personal tastes and crowd wisdom, yield the most significant benefits, outperforming models that solely use either setting $\mathbf{S1}$, $\mathbf{S2}$, or $\mathbf{S3}$.

\section{Discussions: Enhancing Model Performance with More Features}
\label{appendix_C}



In addition to using the 21 instances of tree representations as features (i.e., setting $\mathbf{S4}$), we generate further features associated with users and products. These include the number of ratings given/received by a user/product, the number of active days a user/product gives/receives ratings, whether a user has previously reviewed the same product, and the time interval from the current rating to the most recent and earliest (first) rating by a user or from a product, among others. This process results in a comprehensive set of 136 features, which are then utilized as inputs for LightGBM, RNN, and TabTransformer.

\begin{figure}[h]\small
  \centering
  \includegraphics[width=\columnwidth]{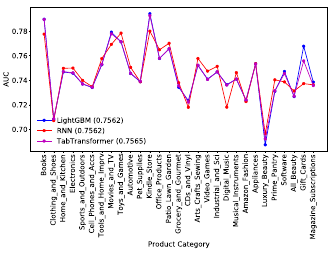}
  \vspace{-0.2in}
  \caption{This figure displays the AUC values of LightGBM, RNN, and TabTransformer with 136 features across 29 individual product categories. The values in parentheses represent the AUC values for the entire product categories.}
  \label{fig:AUC_3_large_models}
\end{figure}

Figure ~\ref{fig:AUC_3_large_models} displays the AUC values obtained from the three models across 29 individual categories. Notably, LightGBM and TabTransformer yield nearly identical AUC values across all categories, with the exception of the ``Gift Card'' category. The RNN model outperforms LightGBM and TabTransformer in some categories, while underperforming in others, even though the overall AUC values of the three models are almost identical. In future work, it would be worthwhile to investigate the causes of such differences among the models. The learning-from-disagreement approach~\cite{wang2022learning,wangJP2022learning} may provide insights into this.

\section{Case Studies}
\label{appendix_D}
\begin{figure*}
  \centering
\includegraphics[width=0.99\linewidth,height=1.0\linewidth]{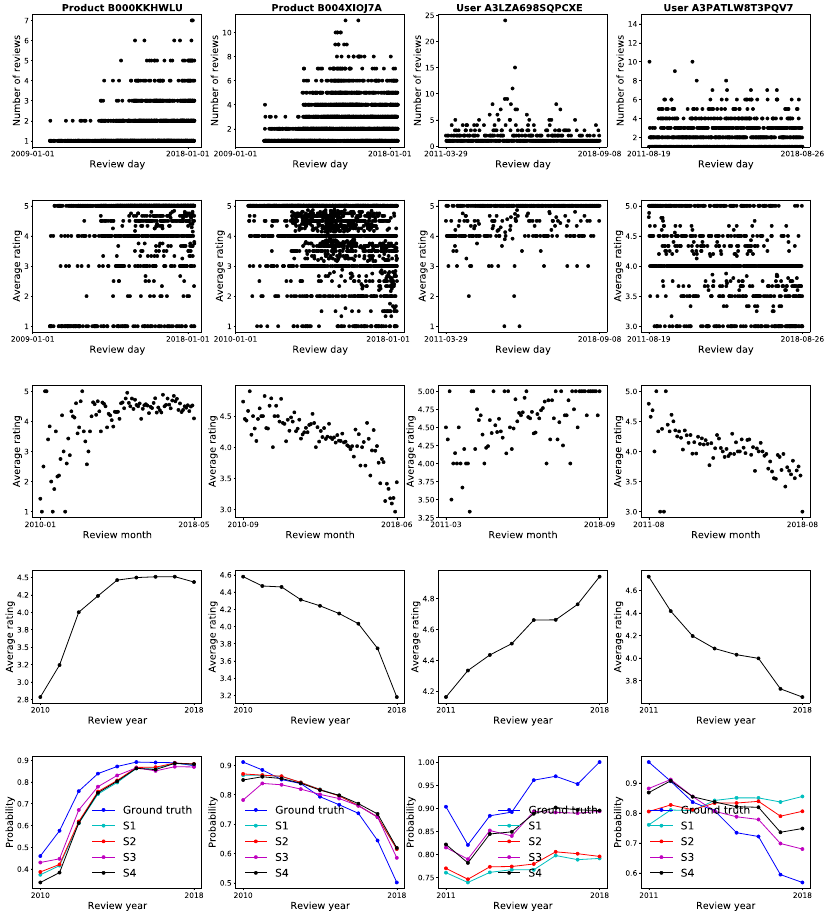}
  \caption{Case Studies: Temporal Dynamics of Two Products and Two Users. The left two columns depict the temporal dynamics of two products, while the right two columns illustrate the behavior of two users. In the top panel, we observe the daily count of ratings. The subsequent rows present average ratings, calculated by dividing the sum of rating values by the count of ratings on a daily, monthly, and yearly basis, respectively. Notable patterns emerge in the monthly and yearly average ratings. The bottom panel displays the predicted probabilities of receiving high ratings from the LightGBM model under four different settings ($\mathbf{S1}$, $\mathbf{S2}$, $\mathbf{S3}$, and $\mathbf{S4}$) compared to the ground truth on a yearly basis. These probabilities are determined as the average scores, while the ground truth is computed by summing the number of high ratings and dividing by the total count of both low and high ratings.}  \label{fig:case_study_full}
\end{figure*}

Figure \ref{fig:case_study_full} provides a visual representation of the evolution of ratings over time for two selected products and two users. The ratings are predicted at the instance level, i.e., for each review, on a daily basis, and then aggregated into yearly ratings to better discern the overall trend. The two panels on the left illustrate the temporal dynamics of the products, while the two panels on the right show the rating behaviors of the users. Product B000KKHWLU is categorized under ``Tools and Home Improvement'', while Product B004XIOJ7A belongs to the ``Electronics'' category. 

The top panel displays the daily count of ratings. Subsequent rows represent average ratings on a daily, monthly, and yearly basis, respectively, based on the aggregation of ratings at the instance level, i.e., for each individual review. Noteworthy patterns emerge in the monthly and yearly average ratings: Product B000KKHWLU (and User A3LZA698SQPCXE) exhibits a clear upward trend in ratings, while Product B004XIOJ7A (and User A3PATLW8T3PQV7) demonstrates a significant downward trend in ratings. This signifies the existence of both positive and negative dynamics within the dataset. The bottom panel presents the predicted probabilities of receiving high ratings from the LightGBM models under four different settings ($\mathbf{S1}$, $\mathbf{S2}$, $\mathbf{S3}$, and $\mathbf{S4}$) in comparison to the ground truth on a yearly basis. All models effectively follow the trend for the two products. However, models with settings $\mathbf{S1}$ and $\mathbf{S2}$, which place importance on crowd wisdom, show significant deviation for the two users, underscoring the vital role of individual user preferences in predicting ratings.


\end{document}